\documentclass[aps,prl,twocolumn]{revtex4}
\usepackage{graphicx}

\begin{document}
\title{Cascaded generation of multiply charged optical vortices
and spatiotemporal helical beams in a  Raman medium}

\author{A.V. Gorbach}
\affiliation{Centre for Photonics and Photonic Materials, Department
of Physics, University of Bath, Bath BA2 7AY, UK}
\author{D.V. Skryabin}
\affiliation{Centre for Photonics and Photonic Materials, Department of Physics,
University of Bath, Bath BA2 7AY, UK}

\begin{abstract}
Using an example of a Raman active medium we
describe how a common nonlinear process of
four-wave mixing can be used to induce strong coupling
between the spatial and temporal degrees of freedom in optical waves.
This coupling produces several unexpected effects. Amongst those are
cascaded excitation of multiply charged optical vortices,
spatial focusing in a nonlinearly defocusing medium and
generation of  helically shaped spatio-temporal optical
solitons.
\end{abstract}

\maketitle

The nonlinearly mediated interaction between frequency harmonics and
the associated effect of frequency conversion have laid the foundation
of nonlinear optics \cite{shen}, and have had an important
effect in  areas such as fluid dynamics and plasma
physics  \cite{fluids}, and - most recently -  physics of
ultracold matter \cite{bec}. Four-wave mixing is the most common
type of nonlinear wave interaction seen in both crystals with
inversion symmetry and  isotropic media. In the context of nonlinear optics the
four-wave mixing process is  typically mediated by
Kerr and Raman nonlinearities  \cite{shen}.

Nonlinearity can be used not only to control the spectral,
and hence the temporal, properties of waves, but also to transform the linear \cite{shen}
and orbital angular momenta of waves \cite{dhol,des,dima,malomed,raman}. In
particular, experiments with the second harmonic generation by
optical vortices \cite{des} have demonstrated that two photons with frequency
$\omega$ and an orbital angular momentum (OAM) quantum number (or vortex
charge) $l$ produce a single photon with frequency $2\omega$ and OAM
quantum number $2l$ \cite{dhol}. Analogous OAM conversion rules have
been reported  for  soliton-like beams within the second harmonic
generation model \cite{des,dima}, for degenerate four-wave mixing
in Kerr-like materials \cite{malomed}, and for  three-wave
Raman resonant process with  higher order Bessel-beams \cite{raman}.
Though the interplay between spatial and temporal dynamics in
nonlinear media has recently become an active research topic, see, e.g. \cite{inter}, it
has failed so far to exploit many of the promising consequences of the
strong spatio-temporal coupling seen in  frequency
conversion experiments using beams with non-zero OAM.

In this work we suggest a method for manipulating the frequency and angular harmonics of
light opening new opportunities for simultaneous spatial and
temporal shaping of optical waves. We will demonstrate that the exploitation of the
coupling between spatial and temporal degrees of freedom
can lead to  such exciting effects as cascaded vortex generation,
strong spatial focusing induced by self-defocusing nonlinearity
and the generation of spatio-temporal helical beams in the solitonic and
non-solitonic regimes.

One of the most efficient nonlinear optical processes
is the Raman mediated four-wave mixing in gases,
which easily produces dozens of spectral lines.
In particular, it has been demonstrated  that the two frequency
excitation of a Raman transition away from the resonance  can
generate, by means of the cascaded four-wave mixing, broad
frequency combs and induce nonlinearity related chirp
\cite{sokolov,korn,sokolov1}. The nonlinear  chirp can be compensated for by
the intrinsic group velocity dispersion (GVD) of the material
resulting in the generation of  trains of ultra-short (close to
single-cycle) pulses \cite{sokolov,korn,sokolov1}. Note, that the
generation of the frequency combs with this technique relies not on
the Raman gain, but on a four-wave mixing process which dominates
the wings of the Raman line and induces strong effective Kerr
nonlinearity \cite{prl}. The starting idea for the results
presented below is that the
cascaded off-resonant Raman process, as in \cite{sokolov}, is
initiated under  conditions where one of the two (frequency detuned) driving fields is  a
singly charged vortex beam.  In this case, the phase dependent nonlinear
coupling between the Raman side-bands (the same coupling, as that which drives
frequency conversion) causes cascaded generation of multiply charged
vortices. A dimensionless model describing  the evolution of the  side-bands is
\cite{sokolov,ol}
\begin{equation}
\label{eqE}
i\partial_{z} E_n-\frac12\Delta E_n=\beta_n E_n + q^* E_{n-1}+ q E_{n+1},
\end{equation}
where $n=-N,\dots,0,\dots, N$ and $\Delta=\partial_x^2+\partial_y^2$.
$E_n$ are the dimensionless amplitudes of the sidebands, such that the total field
is given by $E_{tot}=\sum_n E_n(x,y,z)e^{i\Omega_n t-iK_nz}$,
where $\Omega_n=(\omega_0+n\omega_{mod})/\omega_{mod}$.
$\omega_{mod}=\omega_1-\omega_0$ is the modulation frequency (which is the frequency
difference between the two driving fields) and $n=0,1$.
The physical frequencies and wavenumbers are represented by the lower case letters
$\omega_n$ and $k_n$, whilst their dimensionless equivalents  - by the upper case:
 $\Omega_n$ and $K_n$.
The dimensionless time $t$ is measured in  units of  $1/\omega_{mod}$,
the propagation coordinate $z$ is in  units of $L$, and the
transverse coordinates $(x,y)$ are in  units of $\sqrt{Lc/\omega_0}$.
$K_n=(\omega_0+n\omega_{mod})L/c$ is the scaled free space wavenumber.
Here,  $L=(\eta\hbar\omega_0{\cal N}|b|)^{-1}$ characterizes the coupling length over which power
is transferred between  neighboring side-bands in the absence of dispersion.
$\eta\approx 376$, $\cal N$ is the density of molecules and $b$ is a coefficient
characterizing the material dependent coupling  between the sidebands \cite{sokolov}, which
weak frequency dependence is neglected for simplicity.

\begin{figure}
\includegraphics[width=0.47\textwidth]{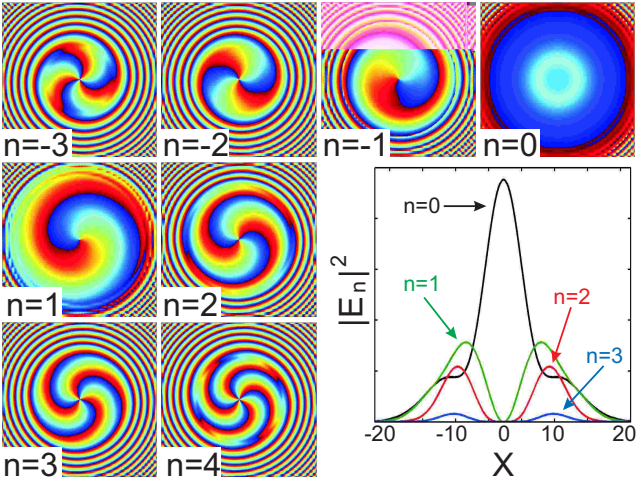}
\caption{(Color online) The smaller panels show the transverse  profiles of the sideband phases
($\arg(E_n(x,y))$). The larger panel shows the $x$-distribution of
intensities, $|E_n(x,y=13)|^2$. The propagation distance is $z=0.2$.
The window size in each plot is $25 \times 25$
dimensionless units (which corresponds to
$\simeq 600 \times 600 \mu m$). The other parameters are:
$\mu=5$, $j=-1$, $p_2=0.005$, $N=23$. The initial conditions are: $A_0=\sqrt{2}$,
$A_1=1$, $w_0=2\sqrt{5}$, $w_1=\sqrt{10}$ (see text for more details).}
\label{fig1}
\end{figure}

$L$ varies from $1$ to a few mm for $D_2$ and $H_2$ gases \cite{sokolov},
so that the one unit of $x$ corresponds to a few tens of microns.
$q$, the Raman coherence, is responsible for the coupling
between the side-bands. In the adiabatic approximation \cite{ol,pra}
\begin{equation}
\label{stacQ}
q={jS \over 2\sqrt{\mu^2+|S|^2}},\quad S=\sum_nE_nE_{n+1}^*,
\end{equation}
where $\mu=|\omega_{mod}-\omega_r|/(|b|I_0)$
is the scaled modulus of the detuning of the modulation
frequency from the Raman frequency $\omega_r$ and $j=sign(\omega_{mod}-\omega_r)$
is the detuning sign. $|q|$ varies from $0$ to $1/2$ for $|S|$ varying from $0$ to $\infty$.
$j=+1$ or $-1$ correspond, respectively, to the self-focusing or self-defocusing nonlinearity
induced by the Raman coherence \cite{prl,ol,pra,exp}.
$E_n\sqrt{I_0}$ are the dimensional amplitudes of the harmonics.
For $D_2$ and $H_2$ gases $\mu=1$ corresponds to  $I_0\sim 0.1$GW/cm$^2$,
provided  $|\omega_{mod}-\omega_r|\sim 1$GHz.
The frequency dependence of the propagation constant $\beta(\omega)$
can be fitted with a polynomial, which in the discretized frequency space gives
$\beta(\omega_n)= \beta_n=\sum_{i=0}^M p_i n^i$. Without any loss of generality the
$p_{0,1}$ coefficients can always be set to zero by phase rotations \cite{ol}.
$p_2$  and $p_{i>2}$ characterize  GVD and  higher order dispersions respectively.

In order to derive the rules for the OAM conversion  we assume that
the system is pumped  with two neighboring sidebands ($n=0,1$)
having phases $\phi_0$ and $\phi_{1}$, where
$\phi_n=\Omega_nt-(K_n-\beta_n)z+l_n\theta$, $\theta=arg(x+iy)$ is
the polar angle and $l_n$ is the OAM quantum number of the n's
sideband.  The conditions for the phase matched  excitation of the
nearest Stokes and anti-Stokes sidebands are
$2\phi_0-\phi_{1}=\phi_{-1}$, $2\phi_1-\phi_0=\phi_2$.  For an arbitrary $n$ these
are generalized as $2\phi_n=\phi_{n+1}+\phi_{n-1}$. While the $z$-dependent
part  of the above condition can not be made zero in  dispersive
materials ($\beta_n\ne 0$), the $t$  and $\theta$  dependent parts
are nulled exactly for $\omega_{n}=\omega_0+n\omega_{mod}$ and
$l_n=l_0+n(l_1-l_0)$. Therefore, the  OAM conversion rule is
analogous to the one for the frequencies and $l_1-l_0$ can be
considered as the seeded modulation of OAM. In a particular example
considered below $l_0=0$ and $l_1=1$, so that $l_n=n$.
\begin{figure}
\includegraphics[width=0.45\textwidth]{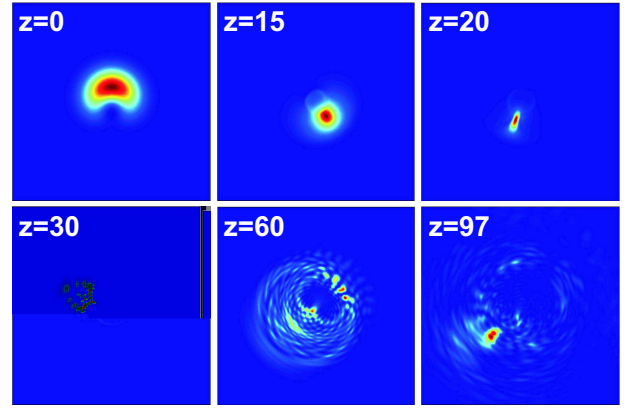}
\caption{(Color online) Intensity distribution
of the total field $|E_{tot}|^2$ after different propagation distances $z$ for $t=0$.
Azimuthal compression is seen for $z=20$ and subsequent recompression for $z=97$.
The nonlinearity is defocusing, $j=-1$. $w_0=20\sqrt{5}$, $w_1=10\sqrt{10}$.
The other parameters are the same as in Fig.~\ref{fig1}.
The window size  is $250 \times 250$ dimensionless units.}
\label{fig2}
\end{figure}
\begin{figure}
\includegraphics[width=0.45\textwidth]{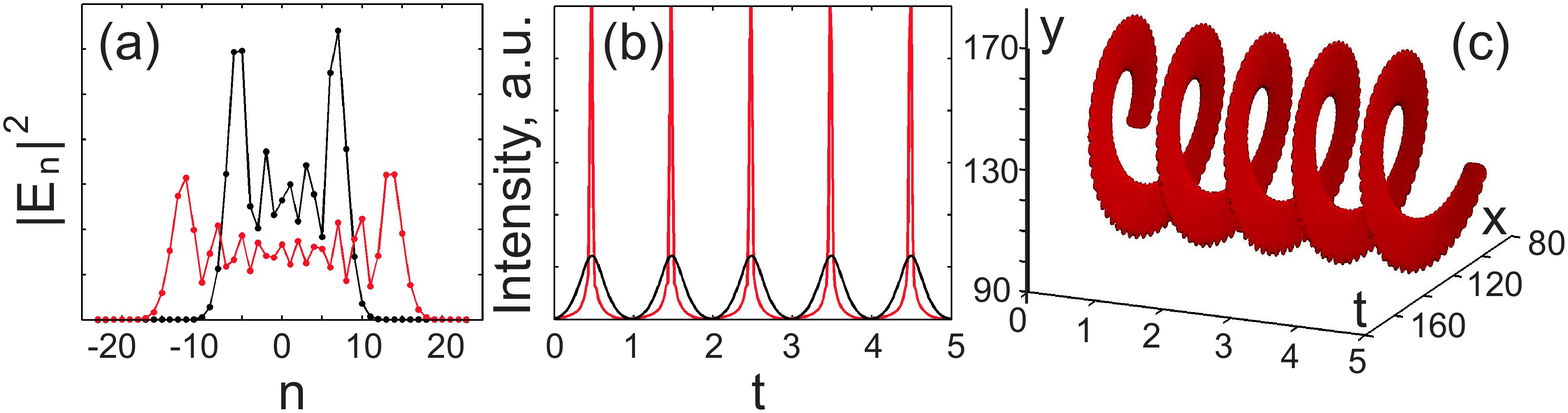}
\caption{(Color online)
 Spectral (a) and temporal (b) profiles of the total field
$z=10$ (black lines) and $z=20$ (red lines).
To plot (a,b) we used the cut through the helix taken
at the point of maximal intensity $|E_{tot}|^2$. Other parameters
are the same as in Fig.~\ref{fig2}. (c) Isointensity $(x,y,t)$-plot
at $50\%$ of maximum for $z=20$.}
\label{fig3}
\end{figure}
For sufficiently strong input fields  Raman four-wave mixing
should develop  in a cascaded manner
and excite higher-order Stokes and anti-Stokes fields
carrying progressively larger OAM quantum numbers. This process is
illustrated in Fig. 1, where we show the transverse profiles of the phases
of the side-bands found by numerical integration of
Eqs. (\ref{eqE}), (\ref{stacQ}) with 47
sidebands and the initial conditions $E_{0}=A_0 e^{-r^2/w_0^2}$,
$E_{1}= A_1 r e^{i\theta-r^2/w_1^2}$,
$E_{n\ne 0,1}=0$ ($r^2=x^2+y^2$). One can see that the generated
Stokes ($n<0$) and  anti-Stokes ($n>0$) components carry progressively
increasing  vortex charges, which can be inferred from the number of jumps
between the blue (zero phase) and red (phase $2\pi$) colors
(see areas around the beam centers in Fig. 1).
The efficiency of the OAM conversion (in terms of the phase matching requirements)
is the same as the efficiency of the frequency conversion. However, when dealing
with  spatially inhomogeneous beams one should provide
sufficient initial overlap of the two pump fields to build a strong coherence $q$.
This is achieved by using a vortex free Gaussian beam  wider than
the one with a vortex ($w_0>w_1$).

The above result suggest that the methods of  frequency domain wave synthesis
\cite{sokolov,korn,sokolov1} have the potential to be extended
to the angular momentum components, and so can be used for simultaneous spatial and temporal beam shaping.
This is because, the frequency harmonics in our scheme
are, at the same time,  the angular harmonics.
To start analyzing problem of  spatio-temporal beam shaping it is instructive
to get some analytical information about the phase evolution under the
condition of  two field excitation. First, we assume that $E_n=f_n(r,z)e^{in\theta}$ and
that  the fields apart from
$n=0,1$  are initially zero. Then by neglecting  the dispersion
($\beta_n=0$) and the $r$-derivatives of $f_n$ (i.e., the radial diffraction),
we can find an explicit analytical solution to
Eqs. (\ref{eqE}), (\ref{stacQ}). This solution gives us $z$-dependence for all
$f_n$'s in terms of Bessel functions \cite{sokolov}. Using known identities
for the Bessel functions one can find  the expression for
the total field
\begin{eqnarray}
\label{etot_spiral}
&& E_{tot}\simeq e^{i\left[\Omega_0 t-iK_0z + 2jz|q_0| \cos(t+
\theta-Kz+\phi_0)\right]}\\
\nonumber && \times
\left[f_{1}(z=0)+f_{0}(z=0)e^{-i(t-K z+\theta)}\right],~K\equiv L\Omega/c.
\end{eqnarray}
where $q_0=q(z=0)$.
In our case, the known result for $E_{tot}$ \cite{sokolov}
is altered  by the presence of $\theta$ inside
both the phase chirp and amplitude modulation. From the amplitude modulation term
one can see that the emerging intensity pattern of $|E_{tot}|$ corresponds to a helix.
The helical shape of the isointensity surfaces can be seen in both the $(x,y,t)$
and $(x,y,z)$ subspaces. The time period of the first is $2\pi$ and
the spatial period of the second is $2\pi/K$.
Changing $l_1$ to $-1$ changes the sign in front of $\theta$.
One can also note, that the directions of rotation
({\em chirality}) of the helices in $t$ and $z$ are opposite.

The phase chirp in (\ref{etot_spiral}) is induced by the Raman coherence
and plays a crucial role in the evolution of the field,
when dispersion and diffraction are taken into account.
In typical cases the background dispersion of the gas is normal,
meaning that the compensation of the Raman chirp occurs for negative
detunings from the Raman resonance ($j=-1$).
Since diffraction in general and, in particular, angular diffraction ($\sim\partial_{\theta}^2$) is
 mathematically equivalent to  anomalous GVD,
it seems at the first glance that the simultaneous temporal
and azimuthal compressions for materials with normal GVD are not possible,
because the coherence induced azimuthal and temporal chirps have the same sign $j$. The
reality, confirmed by numerical modeling, is, however,  more intriguing. Indeed, the angular and
frequency harmonics in our case have the same complex amplitudes $f_n$.
This implies that  strong coupling between spatial and temporal degrees
of freedom occurs in our system. Therefore we can
assert that if   the characteristic length $L_{com}$, over
which maximal pulse compression is achieved,
is shorter than the characteristic diffraction length $L_d$,
then the temporal and azimuthal compression should develop together.
Modeling  Eqs. (\ref{eqE}) with a  plane wave excitation and $q_0 \simeq 0.5$ we
found  that $L_{com}\simeq 20$
(corresponding to a physical distance of $\sim 2-4$cm), which agrees with Refs. \cite{sokolov,sokolov1}.
In our dimensionless notations $L_d$ is readily estimated as $L_d\sim w_{0,1}^2$.

To observe the effect of  simultaneous temporal and azimuthal compression
with $j=-1$ (defocusing nonlinearity) and normal GVD ($p_2>0$)
we take the  initial excitation like in the modeling shown in Fig. 1,
but use wider beams  ($w_0=10\sqrt{20}$, $w_1=10\sqrt{10}$), in order
to minimize the role of diffraction over the propagation distances $\sim L_{com}$.
The results of this modeling are shown in Figs. 2,3. One can see that for a propagation
distance of $z=20$ there are around $30$ harmonics generated, see Fig. 3(a),
implying the presence of  vortex beams with $|l_n|=15$. Simultaneously,
a high degree of the temporal
and azimuthal compression is achieved, cf. Fig. 2 (z=20) and Fig. 3(b).
With further propagation, see Fig. 2, the
pulse starts to spread out again producing complicated spatial patterns.
However,  the compression in this scheme is a quasi-periodic
process, and  for $z=97$ we  observe a clear signature
of the second,  less pronounced, temporal and azimuthal compression.
The spatio-temporal $(x,y,t)$ plot in Fig. 3(c) shows  helical structure of the intensity
$|E_{tot}|^2$ for the propagation distances before strong
defragmentation of the transverse profile begins.
If we change the detuning sign to $j=1$ (focusing nonlinearity),
then the spatiotemporal dynamics is qualitatively different. The
compression effect is absent and with the same initial conditions
we  observed small scale spatial modulational instability and subsequent self-focusing
of the emerging filaments developing at $z>30$.

Self-focusing nonlinearity balanced by  diffraction  suggests the
possibility of spatial solitons. Indeed, the latter have been recently reported
for the simplest case of a two-component Raman model \cite{pra}, while Raman
self-focusing has been seen experimentally \cite{exp}. The question, which
is relevant for our problem, is whether  multi-frequency solitons carrying OAM can be found.
To find these structures we assumed that $E_{n}(x,y,z)=f_n(r)e^{in\theta+i(\kappa_1+n\kappa_2)z}$,
substituted it into Eqs. (\ref{eqE}) and solved the nonlinear
system of ordinary differential equations for $f_n(r)$ numerically.
$\kappa_{1,2}$ are the soliton parameters chosen to ensure decay of the soliton tails
for $r\to\infty$. The boundary conditions used at $r=0$ are
$\lim_{r\to 0}f_nr^{-|n|}=const_n$. Using this approach we have found
a two-parameter ($\kappa_1,\kappa_2$) family of solitons, one example of which
is shown in Fig. 4(a).  The total intensity corresponding to this solution
has a helical profile in both $(x,y,t)$ (not shown, but it is similar to Fig. 3(c))
and $(x,y,z)$ subspaces, see Fig. 4(b). Solitons for sufficiently small values of $\mu$
(implying sufficiently large values of the intensity) are robust
with respect to small perturbations, see Fig. 4(b). Taking larger values of $\mu$
leads to excitation of internal oscillations and wobbling helical trajectories, see Fig. 4(c).
Notably, experimentally realistic excitation, by just two beams with $l_0=0$, $l_1=1$,
leads to the emergence of  structures very close to the found soliton solutions, see Fig. 4(d).
This is providing that the input beams are sufficiently narrow, so that the diffraction length
is short in comparison to the dispersion length and the self-focusing kicks in before
GVD desynchronizes  the phases of the harmonics.

\begin{figure}
\includegraphics[width=0.45\textwidth]{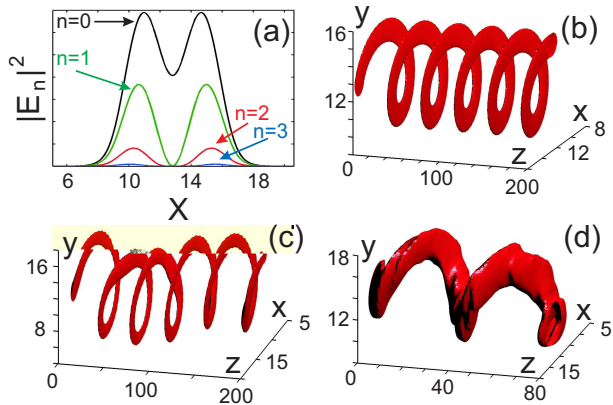}
\caption{(a) Transverse profiles of $|E_n(x,y=13)|^2$ for the multi-frequency soliton with
$\mu=1$, $j=1$, $p_2=0.005$. Soliton parameters are: $\kappa_1=-0.5$, $\kappa_2=0$.
(b) Corresponding evolution of the total intensity
$|E_{tot}|^2$ along $z$ with $1\%$ noise in the initial conditions.
Isointensity plot at $80\%$ of maximum.
(c) The same as (b), but for $\mu=5$. (d)
Isointensity plot at $20\%$ of maximum for the case of the initial excitation with two beams.
$j=1$ and the other parameters as in Fig. 1.
In (b),(c) and (d) the total intensity
is evaluated for $t=0$.}
\label{fig4}
\end{figure}

The helical soliton beams reported here are qualitatively different from the
so-called spiralling solitons or rotating soliton clusters \cite{spir}, which sustain their
rotation due to the interaction between the individual beams accompanied by the conservation of
the angular momentum. In our case the
helical evolution does not require the presence of more than one intensity lobe
and originates from the interaction of multiple frequency harmonics carrying
progressively growing OAM. Ref. \cite{vas} studied complex spatial patterns
emerging from the linearly
interacting vortex beams carrying different OAM, but having  identical frequencies.
Multicomponent spatial solitons carrying OAM have been reported in \cite{segev} for
a  nonlinearity which does not depend on the relative phases of the interacting harmonics,
so that the four-wave mixing mediated interaction of the beams has not been included
(incoherent interaction). In this case, the  OAM of individual components
can be arbitrary, i.e. it is not controlled by any selection rules, and higher
order vortices can not be generated by the system itself.

In summary: We have demonstrated that a cascaded four-wave mixing process
in an off-resonantly excited Raman medium with one of the two input
fields being an optical vortex leads to  generation of
multiply charged optical vortices. Each newly generated vortex
beam has its own frequency creating strong dependence between the spatial and
temporal degrees of freedom, which can be used for new forms of optical wave
synthesis. In particular, we have demonstrated the generation of
simultaneous azimuthally and temporally compressed pulses on the defocusing
side of the resonance and helical optical solitons on the focusing side.
The suggested technique and observed effects pave the way for practical implementations
of  spatial wave shaping using methods of spectral control. The above concepts are also
likely to be  applicable for the  waves of a different nature.

\end{document}